\documentstyle[12pt]{article}
\begin{document}
\def\baselinestretch{1.2}
\hfill{{\tt }
\begin{center}
{\bf A Precise  Determination of the Pion-Nuclear \\
Coupling Parameter from Weak Processes in ${}^{3}\!He$} \\[2mm]
Nimai C. Mukhopadhyay$^{a,b}$ and K.~Junker$^a$ \\[1mm]

{\it
 $^a$ Paul Scherrer Institut, CH--5232 Villigen PSI, Switzerland
} \\ [2mm]

{\it
 $^b$ Physics Department, Rensselaer Polytechnic Institute \\
 Troy, NY 12180-3590} \\[2mm]
\end{center}

\begin{center}
\end{center}
 
\begin{abstract}
We utilize precise weak interaction experiments
on  atomic muon capture and beta decay in the
$A = 3$ nuclei, and take into account
the effects of nuclear ``anomalous thresholds" 
to extract the pseudoscalar $\pi{}-^{3}\!He{}-^{3}\!H$
coupling parameter, $G^{eff}(m_\pi^2) = 45.8 \pm 2.4$. This is
an order of magnitude
improvement in precision over that from the use of pion-nuclear
scattering data and dispersion relations.
\end{abstract}

Weak interaction processes, in which atomic muons get captured by the
nucleus [1], or the NMC, are clean ways to study the semi-leptonic
hadron form factors at low $q^2$ in a nuclear environment of interest to
QCD [2]. These also give important insights into meson exchange
currents (MEC) [3]. The NMC can be, in special cases, useful to give
precise information on the strong pion-nuclear coupling strength, as
we demonstrate below in the $A = 3$ nuclear system.

The process [3-5]

\begin{equation}
   {}^{3}\!He + \mu^{-}(1S)\rightarrow{}^{3}\!H + \nu_{\mu}
\end{equation}
is attractive theoretically for a number of reasons : (a) the weak
hadronic current in (1) has the same Lorentz structure as the fundamental
nucleon process, $p + \mu^{-}(1S)\rightarrow n + \nu_{\mu}$ [1].
(b) The nuclear physics of the $A = 3$ system has been carefully studied.
 Thus, explicit wave functions can be computed with great reliability [3,6].
(c) The MEC contributions [3] can be determined in a parallel fashion to
that of the nuclear $\beta$ - decay [7]:
\begin{equation}
    {}^{3}\!H\rightarrow{}^{3}\!He + e^{-} + \bar {\nu_{e}}.
\end{equation}

Recently an experimental breakthrough has been achieved [5] for the study
of (1) at the ``muon factory" of the Paul Scherrer Institut (PSI). The
atomic boundary conditions in the 2S and 1S states [1] have been
carefully controlled, confirming a statistical hyperfine 
atomic population in the $1S$ state before muon capture by the
${}^{3}\!He$ nucleus. This yields a precise capture rate:
\begin{equation}
      \Lambda_{c} = 1496 \pm 4 s^{-1}.
\end{equation}
This can be nicely understood theoretically in terms of the weak and
electromagnetic form factors that are known in the $A = 3$ nuclei [4]. 
The importance of the MEC
is also demonstrated by the fact that the impulse approximation
[4] yields a rate about 15\% smaller than (3), and the margin is provided
by the MEC [3].

This brings us to the subject of this Letter : use of the newly obtained
precise NMC rate (3) to determine the pion - ${}^{3}\!He - {}^{3}\!H$ coupling
strength at a $q^{2}$ characteristic of the process (1). We shall compare
this to its value from the $\beta$ - decay process (2), and that at the
pion pole as extracted from the strong $\pi^{\pm}{}^{3}\!He$ scattering [8-13].
Contrary to our naive expectations, the present precision of the extracted
coupling strength from the strong processes (Table I) is actually much
worse than that obtained from the NMC and the nuclear beta decay. The
main point of this Letter is the following: Even though this pion-nuclear
coupling strength parameter makes a relatively small contribution to the 
muon capture rate, through the induced pseudoscalar form factor,
the recent PSI experiment on the ${}^{3}\!He$ is so precise that it
can be used to yield a value of this parameter that is 
 not only consistent with its values extracted from the pion-nucleus scattering experiments, {\it but far more accurate}! All we need is the hypothesis of the
partial conservation of the nuclear axial current (nuclear PCAC)
[4,13], to obtain the NMC obervables in
the so-called ``Elementary particle approach" (EPA) [13]. One consequence
of the nuclear PCAC, the Goldberger - Treiman relation (GTR) between the
pion - ${}^{3}\!He - {}^{3}\!H$ coupling parameters, the nuclear $\beta$ - decay
axial form factor and the pion decay constant, will be exploited
in the presence of {\it anomalous thresholds} [8-12]
in the $A = 3$ nuclei. The latter are obviously absent for the nucleon [1],
and its GTR is known to be largely immune from effects of the three-pion
cut [14] or chiral symmetry breaking corrections [15], making it an
excellent test for PCAC. Given the nuclear PCAC and GTR, we shall determine
the pion-nuclear coupling parameter accurately at the pion pole from weak
processes, by a linear extrapolation.  Conversely, by comparing this extracted 
coupling parameter  with that from pion-nuclear scattering 
and a suitable dispersion relation, we can, in effect, test the {\it nuclear}   PCAC and GTR.  
In spite of large structural differences between the nucleon
and the $A = 3$ nuclei (${}^{3}\!He, {}^{3}\!H$), PCAC and GTR may 
work well in both.

  We begin with the nuclear weak hadron current for the process (1). It is
characterized by a Lorentz structure {\it identical} 
to that of the nucleon, since
both the $A = 3$ nucleus and the nucleon are $J^{\pi} = \frac{1}{2}^{+}$,
$I = \frac{1}{2}$ objects. This is exploited in the EPA. 
The hadron current is given by [3]
\begin{equation}
 j^{\mu} = \bar u({k}')[F_{V}\gamma^{\alpha}+iF_{M}\frac{\sigma^{\mu\nu}q_{\nu}}
{2M}+F_{A}\gamma^{\mu}\gamma^{5}+F_{P}\gamma^{5}\frac{q^{\mu}}{m}]u(k),
\end{equation}
with $q^{\mu} = ({k}' - k)^{\mu}$, $m$, the muon mass; $M$ is the mean
nucleon (nuclear) mass, $\bar u({k}')$, $u(k)$ are the 
spin - $\frac{1}{2}$ nucleon (nuclear)
spinors, $F_i$'s are the usual [1] weak form factors.
We assume conserved vector current (CVC) and ignore ''second-class"
terms [1]. The muon capture rate (3) is given by
\begin{equation}
\Lambda = \frac{G^{2}_{F}}{2\pi}{\left|V_{ud}\right|}^{2}{{N}'}^{2}C
{\left|\varphi_{\mu}(0)\right|}^{2}{\nu}^{2}(1-\frac{\nu}{\sqrt{s}}){G_0^2},
\end{equation}
where we follow the notation of Congleton and Fearing (CF) [4]. The weak
interaction physics from nuclei is contained in the effective coupling
constant squared $G_0^2$ :
\begin{equation}
 G_0^2 = G_{V}^2 + 2G_{A}^2 + (G_{A} - G_{P})^{2}.
\end{equation}
The effective vector, axial vector and pseudoscalar coupling combinations
in terms of the $F_i$'s in (4)
are standard [1,4]. Our interest first lies in the determination of $G_{P}$
from (3), fixing $G_{V}$ and $G_{A}$ from experiment, and translating it into
$F_{P}$. From this we shall extract the $\pi - {}^{3}\!He - {}^{3}\!H$ coupling
parameter for the NMC. The GTR will give us this parameter from the nuclear
$\beta$ - decay. From these two kinematic points, we shall extrapolate it
to the pion pole.

  To proceed further, we write the dispersion relation [9] by exploiting
the PCAC {\it Ansatz} that the divergence $\it D(t)$ of the hadronic axial-vector
current is proportional to the pion field :
\begin{eqnarray}
\it D(t)& = &\left[2MF_{A}+\frac{t}{m}F_{p}\right] \nonumber\\
        & = &-\frac{\sqrt{2}f_{\pi}m_{\pi}^{2}G(t)}{t - m_{\pi}^2} +
        \frac{1}{\pi}\int_{a}^{+\infty} d{{t}'}\frac{Im(\it D({t}'))}
            {{t}' - t}.
\end{eqnarray}
For the muon capture reaction (1), $t\approx-0.96
m^2$,  $f_{\pi}$ is the pion decay constant
($130.8{\pm}0.3$ MeV [15]); $G(t)$ is the pion-${}^{3}\!He-{}^{3}\!H$
pseudoscalar coupling parameter, related to the pseudovector one by the 
usual [11] way.
The integration threshold a above is set by the so-called ``anomalous" 
cuts [8,9,11] beginning at
$t = (1.8m_{\pi})^2$ and $(2.1m_{\pi})^2$, coming from
the deuteron - nucleon  and three-nucleon breakup thresholds 
repectively. {\it These are new features of the
nuclear process (1)}, compared to the NMC by the proton. 
Letting $t = 0$, and ignoring the
contribution from all cuts, we get the standard GTR :
\begin{equation}
     f_{\pi} = \sqrt{2}M\frac{F_{A}}{G(0)}.
\end{equation}

At this stage, we recall the nucleon case of the GTR. There are no anomalous
cuts here. Estimating the three-pion cut following Wolfenstein [14], its effect
is found to be small. Thus, the relation (8) can be used to estimate the value
of $G(0)$. Taking $G(0)\approx{G({m_{\pi}}^2)}$ by PCAC, $t = m_{\pi}^2$
corresponding to the physical pion pole, using the neutron $\beta$ - decay
value of $F_{A}(0) = 1.2601\pm0.0025$ [16], and 
$(G(m_{\pi}^2))^2/4\pi = 14.28\pm0.36$
 [17], the GTR is found to be fulfilled within 5\%. Using (7), one can then
estimate $F_{P} = 7.25\pm0.09$ for muon capture by protons, for which
$t\approx-0.88m_{\pi}^2$. At present, this prediction for the nucleon 
is poorly tested through NMC [1]. In radiative muon capture
(RMC),  there is a  disagreement
with it in a pioneering RMC experiment at TRIUMF [18].

Returning to the nuclear system $A = 3$, let us rewrite the dispersion
relation (7) as
\begin{equation}
 \it D(t) = -\frac{\sqrt{2}f_{\pi}m_{\pi}^2G({m_\pi}^2)}{t - m_{\pi}^2}(1 + \delta(t)),
\end{equation}
where $\delta(t)$ is the nuclear correction from the anomalous cuts. Thus, we
can introduce an {\it effective} pion - nuclear coupling parameter
$G^{eff}(t)$ by the relation [9,11],
\begin{equation}
 G^{eff}(t) = G({m_\pi}^2)(1 + \delta(t)).
\end{equation}
This yields for the nuclear $\beta$ - decay an effective GTR that takes
implicitly into account the effects of the nuclear anomalous cuts:
\begin{equation}
 G^{eff}(0) = \frac{\sqrt{2}MF_{A}(0)}{f_{\pi}},
\end {equation}
by substituting (10) in (9) and taking the $t\rightarrow0$ limits, where
$M$ is the mean ${}^{3}\!He - {}^{3}\!H$ mass, $M\approx2808.7$ MeV and
$F_{A}(0)$ is obtained from ${}^{3}\!H$ $\beta$ - decay [7] :
\begin{equation}
   F_{A}(0) = 1.212\pm0.004.
\end{equation}
This gives, using (12) in (11),
\begin{equation}
 {G^{eff}}_{\pi{}^{3}\!He{}^{3}\!H}(0) = 36.81\pm0.15.
\end{equation}
Note that this extraction {\it does not} require our explicit knowledge of
the anomalous cut contribution, $\delta(0)$.

We now discuss the NMC reaction (1) and see how the latest precision
measurement of $\Lambda_{c}$ yields a value of
${G^{eff}}_{\pi{}^{3}\!He{}^{3}\!H}(t_{cap})$, where $t_{cap}$ is the
characteristic value of $t$ in the capture process (1). Using Eqs. (7), (9)
and (10), we get a nuclear PCAC equation, implicitly including the effects
of anomalous cuts:
\begin{equation}
{G^{eff}}_{\pi{}^{3}\!He{}^{3}\!H}(t_{cap}) = -\frac{(t_{cap} - m_{\pi}^2)}
{\sqrt{2}f_{\pi}m_{\pi}^2}\left[2MF_{A}(t_{cap}) +\frac{ t_{cap}}{m}F_{P}(t_
{cap})\right].
\end{equation}
Using the newly measured rate (3), we can determine a range of the
{\it least known} weak nuclear form factor $F_{P}(t_{cap})$, 
holding the others to
their known values [4]. To do this, we exploit the experimentally known
vector form factors and take the t dependence of $F_{A}$ from the vector one
[4], as is conventionally done. (Future neutrino experiments at facilities like
KARMEN [19] would eliminate this approximation.) This yields, using (3), (5)
and (6),
\begin{equation}
F_{P} = 20.80\pm1.6,
\end{equation}
with the parameter $C$, the correction factor in (5) due to 
the nuclear finite size
effect taken  to be $0.98$. The nuclear PCAC equation (14)
gives us
\begin{equation}
{G^{eff}}_{\pi{}^{3}\!He{}^{3}\!H}(-0.954m^2) = 31.9\pm1.3.
\end{equation}
Eqs. (13) and (16) are two crucial results of this Letter. The effects of
anomalous cuts in the $A = 3$ nuclei are $implicitly$ $included$ in these
numerical values.

   Before discussing the significance of these results, we come to the
determination of the $\pi-{}^{3}\!He-{}^{3}\!H$ coupling parameter from the
strong interaction process in the $A = 3$ system directly. The principle has
been reviewed by Ericson and Locher long ago [8]. One writes down a dispersion
relation for the amplitude, antisymmetric under crossing [8] :
\begin{equation}
 Ref^{-}(\omega) = \sum_{i}\frac{2\omega r_{i}}{\omega^2 - \omega_{i}^2}
 + \frac{2\omega}{\pi}P\int d{{\omega}'}\frac{Imf^{-}({\omega}')}
 {{\omega}'^2 - {\omega}^2},
\end{equation}
where $\omega$ is the pion lab energy, the poles come from the neighboring
nuclei and $r_{i}$ is the residue of the $i$-$th$ pole, the coupling constant
of interest. Several authors [10-12] have made use of the
$\pi^{\pm}{}^{3}\!He$ total cross section data in the physical region and
analytic extraplolation in the unphysical region, using relation (17), wherein
the sum over $i$ gets replaced by a single term, the effective
residue at the pion pole, earlier 
denoted by us as $({G^{eff}}_{\pi{}^{3}\!He{}^{3}\!H})^2$. 
The results of these authors 
yield  a broad range of values and are summarized in Table I, along
with the values obtained from the weak interaction processes (1) and (2). An
important point to note here is {\it the large errors associated with the strong
interaction determinations} of the ${G^{eff}}_{\pi{}^{3}\!He{}^{3}\!H}$
parameter, compared with the precision at the weak interaction values of $t$,
$t\approx0$ and $t = t_{cap}$ for the $\beta$ decay and muon capture
respectively.

Let us now return to the significance of the determination of the
$G^{eff}$ from the weak interaction in (13) and (16) and their implications
at the pion pole. Direct investigations of the effects of anomalous cuts
have been made by Jarlskog and Yndurain [9] and Kopeliovich [11]. They both find
significant variations between $t = 0$ and $t = t_{cap}$ in the value
of $G^{eff}$, due to presence of these cuts. Thus, 
\begin{equation}
\global\def\theequation{18a}
{G^{eff}}_{\pi{}^{3}\!He{}^{3}\!H}(0)
\approx1.09{G^{eff}}_{\pi{}^{3}\!He{}^{3}\!H}(t_{cap}),
\end{equation}
according to Jarlskog and Yndurain, and
\begin{equation}
\global\def\theequation{18b}
{G^{eff}}_{\pi{}^{3}\!He{}^{3}\!H}(0)
\approx1.19{G^{eff}}_{\pi{}^{3}\!He{}^{3}\!H}(t_{cap}),
\end{equation}
according to Kopeliovich. We find, from (13) and (16),
\begin{equation}
\global\def\theequation{\arabic{equation}}
\setcounter{equation}{19}
{G^{eff}}_{\pi{}^{3}\!He{}^{3}\!H}(0) = (1.15\pm0.05)
{G^{eff}}_{\pi{}^{3}\!He{}^{3}\!H}(t_{cap}),
\end{equation}
in qualitative agreement with both theoretical estimates (18a,b), but unable
to distinguish between them. However, {\it the
deviation from unity in the value of the numerical coefficient on the right-hand
side of Eq.(19) is a confirmation, from the weak
interaction experiments, of the role of the anomalous cuts in the
$\pi-{}^{3}\!He{}-^{3}\!H$ coupling.}

    We can now use our pion-nuclear coupling values, obtained from the weak
interaction studies to extrapolate to the pion pole. With 
a linear extrapolation [9],
\begin{equation}
G(t_{cap}) = G(0) + \frac{t_{cap}}{m_{\pi}^2}(G(m_{\pi}^2) - G(0)),
\end{equation}
we get
the $\pi-{}^{3}\!He{}-^{3}\!H$ coupling constant at the pion pole:
\begin{equation}
{G^{eff}}_{\pi{}^{3}\!He{}^{3}\!H}(m_{\pi}^2) = 45.8\pm2.4,
\end{equation}
consistent with the numbers obtained from the $\pi^{\pm}{}^{3}\!He$
scattering (Table I), but far more accurate! We have here 
achieved an improvement in
precision of the determination of the strong pion-nuclear coupling by an
order of magnitude, compared with the current accuracy of 
its inference from the pion-nuclear scattering. This extraction
of a precise pion-nuclear coupling parameter from the weak interaction
processes is the {\it central result} of this Letter.

Further theoretical studies are needed to understand the
dynamical significance of the value of the coupling constant in Eq. 
(21). This much is already clear: {\it The square of the coupling constant
obtained above is about 30\% bigger than the impulse approximation estimate
of Ericson and Locher [8].} 

   In summary, we have studied here the weak interaction observables, the
nuclear $\beta$ - decay rate of ${}^{3}\!H$ to ${}^{3}\!He$, and the inverse
muon capture rate, recently measured at PSI with a great precision, and
used them to determine the pion-nuclear coupling parameter. 
We have utilized the nuclear PCAC and
Goldberger - Treiman relation, taking  the effects of nuclear
breakup channels in the intermediate state, through the anomalous cuts, into
account.
We have extrapolated the
$\pi{}-^{3}\!He{}-^{3}\!H$ coupling parameter from the weak interaction
kinematics to the pion pole and extracted the pion-nuclear coupling constant.
The resultant parameter, $45.8\pm2.4$, is much more precise than 
the values obtained
from the $\pi^{\pm}{}^{3}\!He$ dispersion relations (Table I). Its dynamical
implications need further theoretical exploration beyond the impulse
approximation. Conversely, its consistency
with the values from the existing pion-nuclear scattering analyses 
is a  {\it new
proof} of the validity of  {\it nuclear} PCAC and Goldberger-Treiman relation.
This test could be considerably strengthened with new high-quality experiments
on the pion-nuclear scattering such that the resultant error on the coupling
parameter, inferred from such experiments, 
would be at least of the same order of magnitude as that in (21). 

   Further improvement in precision of the pion-nuclear
parameter is expected, when the new  experimental studies [20] on polarization
observables in the NMC  are finished. 
The polarization observables are far
more sensitive [4,21] to the pseudoscalar coupling $F_{P}$, hence to the
pion-nuclear coupling parameter, than the rates of the nuclear muon capture.
Thus, we are going to obtain informations on strong
interaction physics, from the on-going weak interaction studies, at a level
of precision  {\it even higher} than what we have reported here. 
Therefore, it would be 
useful to have a corresponding improvement in  accuracy in the 
application of the
dispersion relations to the pion-nuclear scattering. This would need new
precise experiments at the pion factories on the total $\pi{}^{3}\!He$
scattering cross sections. For this, the current data base is not 
good enough.

   One of us (NCM) is grateful to Milan Locher for his kind hospitality
at PSI, his interest and many helpful suggestions. We thank
him and Roland Rosenfelder for their critical readings of the
manuscript. We also thank
Jules Deutsch, Valery Markushin and Claude Petitjean for many helpful
discussions on the PSI
${}^{3}\!He$ muon capture experiment.
The research of NCM is supported in part by the U.S. Dept. of Energy.

\newpage
Table I : A comparison of the effective $\pi-^3He-^3H$ coupling parameter
obtained from different processes: the strong $\pi^{\pm}-^3He$ scattering
and dispersion relation (first column), the Goldberger-Treiman relation
and weak $\beta$-decay (second column) and via the pseudoscalar coupling
from nuclear muon capture (third column). References, from which the numbers
in the first column have been extracted are explicitly given. The second
and third columns contain results of this work, yielding a coupling parameter
of $45.8 \pm2.4$ at the pion pole, to be compared with entries in 
the first column. 

\begin{table}[hb]
\begin{center}
\begin{tabular}{ l l l }
\hline\hline
 G from & G from & G from \\
 $\pi^{\pm}{}^{3}\!He$ scatt. & $\beta$ - decay & muon capture \\ \hline
 $38\pm16$ (Spencer[10]) & $36.8\pm0.2$ & $31.9\pm1.3$ \\
 $45\pm19$ (Mach and Nichitiu [12]) & \\
 $49\pm14$ (Nichitiu and Sapozhnikov [12]) & \\
 $57\pm13$  (Kopeliovich [11]) \\
\\ \hline\hline
\end{tabular}
\end{center}
\end{table}
\end{document}